\documentstyle[preprint,prl,aps,psfig]{revtex}

\def\alam{\alpha_{\Lambda}}
\def\alamb{\alpha_{\overline{\Lambda}}}
\def\Lamz{\Lambda^0}
\def\Lamb{\overline{\Lambda}^0}
\def\Lam{\Lambda}

\def\Xim{\Xi^-}
\def\Xib{\overline{\Xi}^+}
\def\be{\begin{equation}}
\def\ee{\end{equation}}
\draft          
\begin{document}

\preprint{LBNL-46257}

\title{
Search for Direct CP Violation in Non-Leptonic Decays of 
Charged $\Xi$ and $\Lambda$ Hyperons
}

\author{
K.B.~Luk,$^{1,2}$
H.T.~Diehl,$^{6,}$\cite{fnal} 
J.~Duryea,$^{5,}$\cite{ucsf}
G.~Guglielmo,$^{5,}$\cite{fnal} 
K.~Heller,$^5$
P.M.~Ho,$^{1,}$\cite{pmho}
C.~James,$^3$
K.~Johns,$^{5,}$\cite{az} 
M.J.~Longo,$^4$
R.~Rameika,$^3$
S.~Teige,$^{6,}$\cite{ind} 
G.B.~Thomson$^6$ \\
\begin{center} {Fermilab E756 Collaboration} \end{center}
}

\address{
$^1$ Physics Division, Lawrence Berkeley National Laboratory, 
     University of California, Berkeley, CA 94720\\
$^2$ Department of Physics, University of California, Berkeley, 
     CA 94720\\
$^3$ Fermilab, Batavia, IL 60510\\
$^4$ Department of Physics, University of Michigan, Ann Arbor, MI 48109\\
$^5$ School of Physics, University of Minnesota, Minneapolis, MN 55455\\
$^6$ Department of Physics and Astronomy, Rutgers--The State University, 
     Piscataway, NJ 08854
}

\date{\today}

\maketitle

\begin{abstract}

A search for direct CP violation in the non-leptonic decays of 
hyperons has been performed.  
In comparing the product of the decay parameters, 
$\alpha_{\Xi}\alpha_{\Lambda}$, in terms of an asymmetry parameter, 
$A_{\Xi\Lambda}$, between hyperons and anti-hyperons in the 
charged $\Xi \rightarrow \Lambda \pi$ and $\Lambda \rightarrow p \pi$ decay 
sequence, we found no evidence of direct CP violations. 
The parameter $A_{\Xi\Lambda}$ was measured to be $0.012 \pm 0.014$.

\end{abstract}

\vspace{0.2in}
\pacs{PACS numbers: 11.30.Er, 13.30.Eg, 14.20.Jn, 14.65.Bt}

\pagebreak

A few years after the discovery of charge-conjugation/parity (CP) violation 
in the neutral-kaon decay\cite{Cronin}, 
Sakharov suggested that this CP asymmetry was one of the three   
conditions necessary for explaining the domination of matter over anti-matter 
in the Universe\cite{Sakharov}.
To date, CP nonconservation is only seen in 
K$^{0}_{\scriptsize\textrm{L}}$ decays, and the 
origin of this phenomenon remains a mystery. 

In 1958, Okubo pointed out that time reversal (T) invariance, or CP 
symmetry under CPT conservation, could be tested by establishing the 
equality of the partial decay rates between $\Sigma^{+}$ and its 
charge-conjugate decay\cite{Okubo}.
Pais independently stressed that if CP symmetry is exact the slope parameter,
$\alam$, of the $\Lamz \to p \pi^{-}$ decay should be equal in magnitude but 
opposite in sign to $\alamb$ of 
the $\Lamb \to \overline{p} \pi^{+}$ decay\cite{Pais}.
However, quantitative analysis on the validity of CP conservation in 
non-leptonic hyperon decays was not available until the early 1980's.
Contrary to the CP asymmetry of K$^{0}_{\scriptsize\textrm{L}}$ 
observed in 1964, which is related 
to K$^{0}$-$\overline\textrm{K}^{0}$ mixing and is called indirect CP violation,
CP nonconservation in the strange-baryon sector is classified as direct CP 
violation and is due to different dynamics  
in the decay of a hyperon and its antiparticle.
Models other than the superweak type~\cite{Wolfenstein} generally predict 
that CP symmetry is broken in strange-baryon decays
\cite{Chau,Brown,Donoghue1,Donoghue2,Iqbal,He}.
Recasting Pais's proposal an asymmetry $A_{\Lambda}$ is defined as 
\be
A_{\Lambda} = \frac {\alam+\alamb}{\alam-\alamb} \ 
\ee
for the $\Lambda$ decay.
The amount of CP-odd effect is found to depend on the strong phase shifts 
of the final state of the decay and the CP violating weak phases which are  
model-dependent.
$A_{\Lambda}$ is estimated to be
$(2-5) \times 10^{-5}$ in the standard model\cite{Donoghue2,He}, 
but it can be as large as a few times $10^{-4}$ in the other models  
\cite{He2,Chang,He3}.
For the charged $\Xi \to \Lambda \pi$ decay, $A_{\Xi}$ is expected to be 
smaller than $A_{\Lambda}$ by about a factor of ten because the strong phase 
shifts of the $\Lambda\pi$ final state are predicted to be small
\cite{Phases}.      

There have been three experimental searches for CP violation in $\Lambda$
decay reported\cite{R608,DM2,PS185}.
The most precise result came from PS185 at LEAR with 
$A_\Lambda$ = $-0.013\pm0.022$.
There is no measurement available for $A_\Xi$.

In this letter we present the result on a new search for direct CP violation 
in hyperon decay by determining the sum of $A_{\Lambda}$ and $A_\Xi$.
In our experiment, E756, the search was performed with polarized 
$\Lamz$($\Lamb$) obtained from the decay of polarized 
$\Xi^{-}$($\overline{\Xi}^{+}$).
According to the Lee-Yang formula, the polarization of the daughter $\Lambda$, 
${\mathbf{P}}_{\Lam}$, in the $\Lambda$ rest frame is related to the 
polarization of $\Xi$, ${\mathbf{P}}_{\Xi}$, in its rest frame by~\cite{PDG86}
\be
{\mathbf{P}}_{\Lam}
 = \frac{(\alpha_{\Xi} + {\mathbf{P}}_{\Xi} \cdot \hat{\mathbf{p}}_{\Lam})
\hat{\mathbf{p}}_{\Lam}
+ \beta_{\Xi} {\mathbf{P}}_{\Xi} \times \hat{\mathbf{p}}_{\Lam}
+ \gamma_{\Xi} \hat{\mathbf{p}}_{\Lam} \times ({\mathbf{P}}_{\Xi} \times
\hat{\mathbf{p}}_{\Lam})}
{\big( 1 + \alpha_{\Xi} {\mathbf{P}}_{\Xi} \cdot \hat{\mathbf{p}}_{\Lam} \big)}
\ ,
\label{eq:polar}
\ee
where $\hat{\mathbf{p}}_{\Lam}$ is the momentum unit vector of the $\Lam$ 
in the $\Xi$ rest frame, and $\beta_{\Xi}$ and $\gamma_{\Xi}$ are the other 
two decay parameters for the $\Xi \to \Lam \pi$ decay.
The distribution of the protons in the $\Lam$ helicity frame, after integrating over the solid angle of $\Lam$ in the $\Xi$ rest frame and the azimuthal angle 
of the proton in the $\Lam$ helicity frame, is given by
\be
\frac{dn}{d \cos{\theta_{p\Lam}}} =\frac{1}{2} \big(1+\alpha_{\Lam}
\alpha_{\Xi} \cos \theta_{p\Lam} \big),
\label{eq:helicity}
\ee
with $\theta_{p\Lam}$ being the angle between the momentum of the proton and
$\hat{\mathbf{p}}_{\Lam}$.
If CP is an exact symmetry, the product $\alpha_{\Lam}\alpha_{\Xi}$ should 
equal $\alpha_{\overline{\Lam}}\alpha_{\overline{\Xi}}$.
By introducing an asymmetry parameter
\be
A_{\Xi \Lam} = \frac{\alpha_{\Lam} \alpha_{\Xi}-
\alpha_{\overline\Lam} \alpha_{\overline\Xi}}
{\alpha_{\Lam} \alpha_{\Xi}+\alpha_{\overline\Lam} \alpha_{\overline\Xi}}
\simeq A_{\Xi}+A_{\Lam} \ ,
\label{eq:alfa}
\ee
CP symmetry can be studied in the $\Xi \to \Lambda \pi, 
\Lambda \to p \pi$ decay sequence.
A non-zero value for $A_{\Xi \Lam}$ will signal the breaking of CP invariance 
in the decay.

\begin{figure}[hbt]
\centerline{\psfig{figure=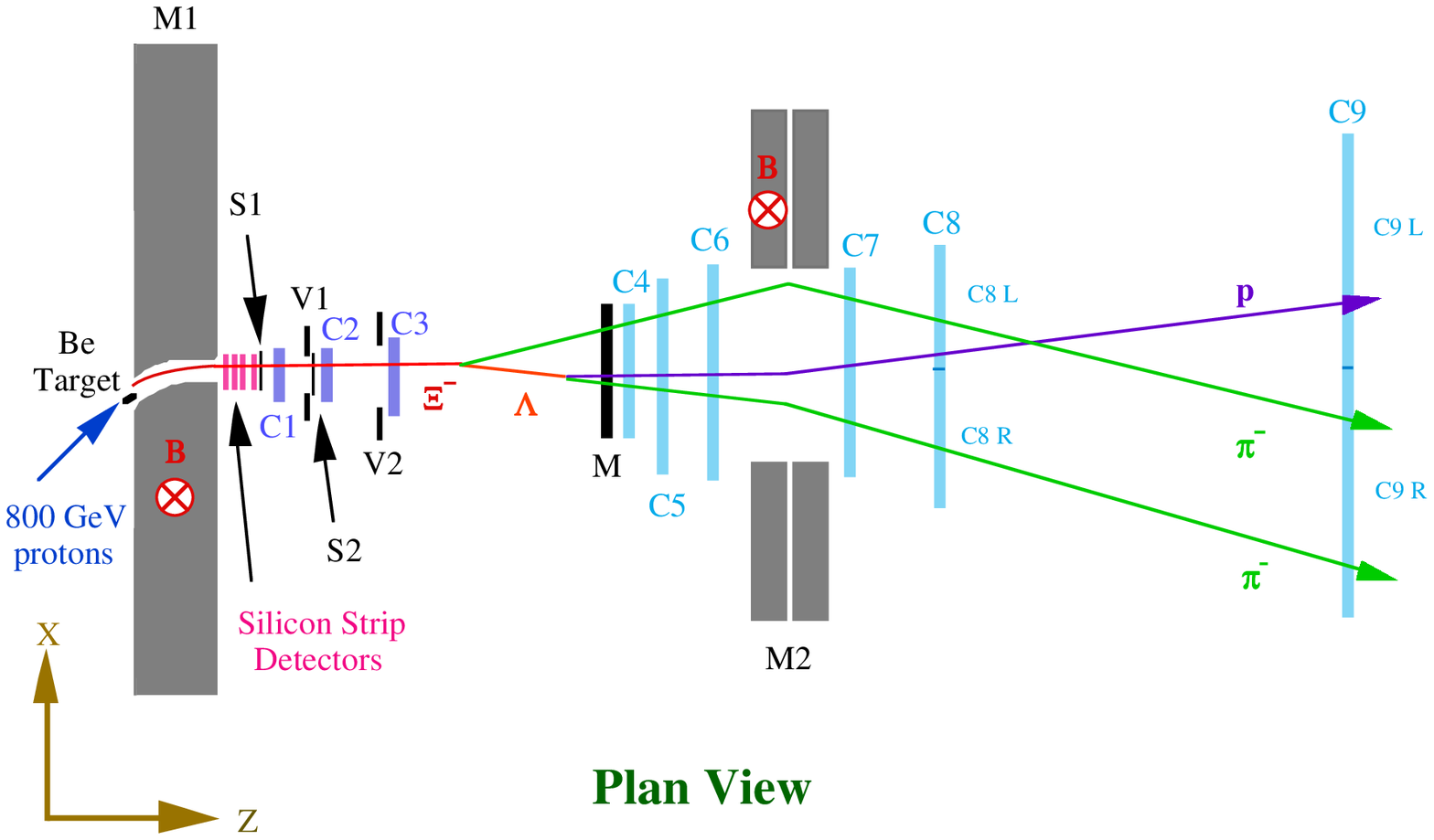,width=6in}}
\caption{Plan view of the E756 spectrometer (not to scale). The x dimensions  
of the silicon strip detectors and C9 are 3 cm and 1.2 m respectively. C9 is 
located at 62.3 m from the exit of the collimator through M1.}
\label{spectrometer}
\end{figure}

Our experiment was carried out in the Proton Center beam line at Fermilab.
Fig.~\ref{spectrometer} shows the plan view of the spectrometer.
The details of the experiment can be found in~\cite{E756} and references 
therein.
An 800 GeV proton beam, with a typical intensity of $~3 \times 10^{10}$ protons 
in 23 sec, was used to produce $\Xi^{-}$ by striking a 
0.2 cm $\times$ 0.2 cm $\times$ 9.2 cm-long 
beryllium target at an angle of 2.4 mrad in the vertical plane relative 
to the proton beam.
The sign of the production angle was flipped regularly to minimize temporal 
systematic problems.
The $\Xi^{-}$ hyperons were momentum selected by a curved channel inside  
a 7.32 m-long dipole magnet M1.
The data presented here were collected with M1 operating at a vertical field 
of 2.09 T.
Typically the rate of the secondary beam was on the order of 100 kHz.
The momenta of the proton and the $\pi$'s from the decays of the $\Xi^-$'s and
$\Lamz$ were measured with eight planes of silicon strip detectors arranged in 
vertical and horizontal views, 
nine multiwire proportional chambers with wire 
spacing of 1 mm (C1, C2, and C3) and 2 mm (C4 to C9), 
and two dipole magnets, M2, 
which deflected charged particles in the horizontal plane with a total 
transverse-momentum kick of 1.5 GeV/c. 
The trigger for detecting the $\Xi^{-} \to \Lambda \pi^{-}, 
\Lambda \to p \pi^{-}$ decay sequence required no hit in V1 and V2, 
hits in both S1 and S2, an analog signal from the multiplicity counter M 
corresponding to at least two but less than five minimum ionizing charged 
particles and a digital signal from the pion side of C8, C8R,  
and one from the proton side of C9, C9L.
In some portion of the data collection, the fields of the momentum 
analyzing magnets were reversed, and the trigger sides of C8 and C9 
were switched to C8L and C9R to improve our understanding of systematics.

To collect $\Xib$ events, the incident proton intensity was reduced to 
an average of about $1 \times 10^{10}$ protons per spill so that the 
secondary-beam intensity did not vary significantly from the negative mode. 
The production angles remained unchanged and were cycled between +2.4 and 
-2.4 mrad.
The polarities of M1 and M2 were reversed and there was no change in the 
triggers.
Hence the experiment was CP invariant to first order.
This greatly reduced the number of potential sources of systematic bias 
due to changes in the spectrometer between the $\Xim$ and $\Xib$ runs.

\begin{figure}[hbt]
\centerline{\psfig{figure=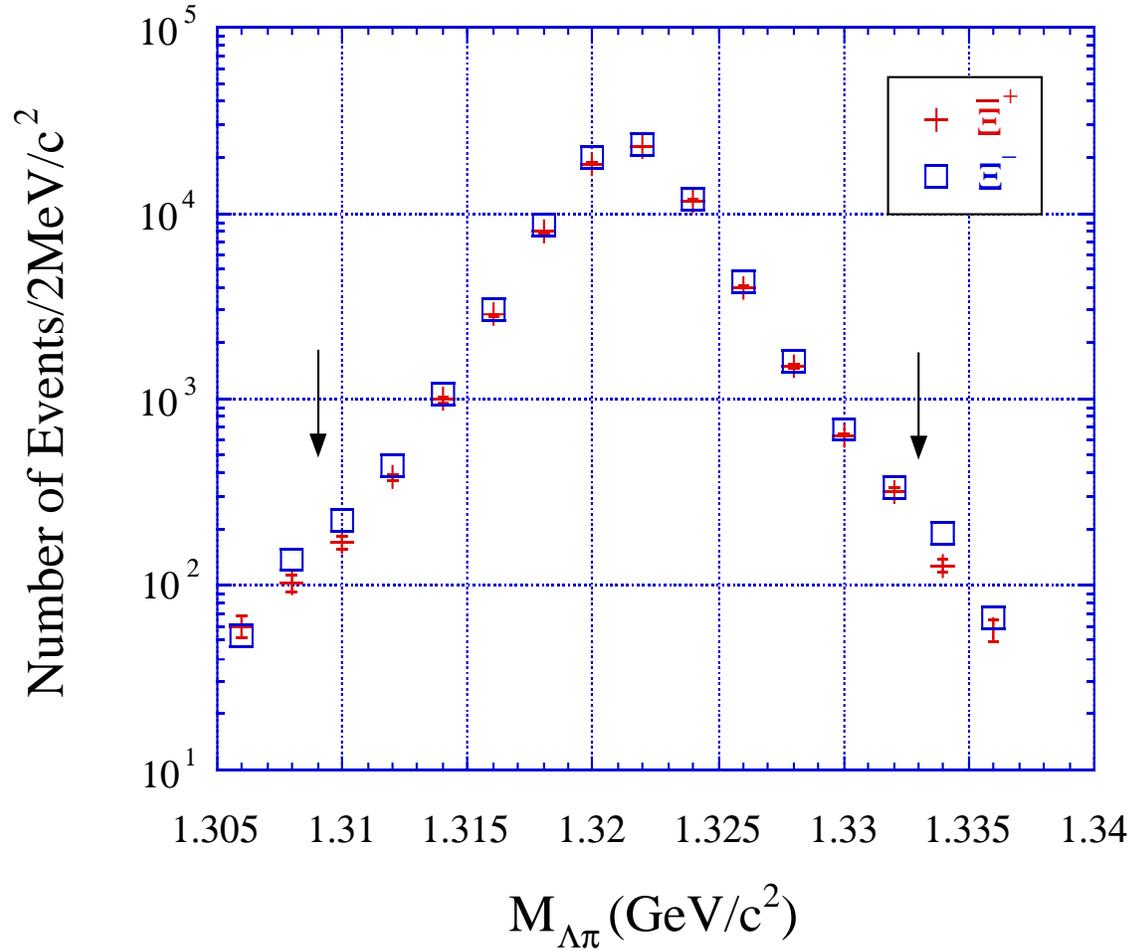,width=6in}}
\caption{Distributions of $\Lambda\pi$ invariant mass with all 
event-selection requirements applied except the cut on $\Lambda\pi$ invariant 
mass. Events between the arrows were used for analysis.}
\label{masses}
\end{figure}

In the offline analysis, data taken with the positively and negatively charged 
secondary beams were processed 
with the same reconstruction program and subjected to identical event-selection 
criteria.
By imposing geometric and kinematic requirements, 
events that satisfied the three-track two-vertex topology were searched for.
The geometric $\chi^2$ for the topological fit of the selected events was 
required to be less than 70 for a mean of 30 degrees of freedom. 
The tracks assigned to be a proton and a pion had to have a p$\pi$ 
invariant mass between 1.108 and 1.124 GeV/$c^2$.
The momentum of the reconstructed $\Xi$ candidate was required to be between 
240 GeV/c and 500 GeV/c, and the track had to  
trace back to within 0.63 cm from the center of the beryllium target in the 
plane normal to the length of the target.
The decay vertex of $\Xi$ was required to be within the fiducial region 
between the exit of the channel, $z$ = 0.25 m, and $z$ = 23 m.
To suppress charged K $\to 3 \pi$ background, the event was also 
reconstructed under the 3$\pi$ hypothesis. 
The resulting 3$\pi$ invariant mass was then required to be greater than 
0.51 GeV/$c^2$.
The comparison of the $\Lambda\pi$ invariant mass distributions between the 
$\Xim$ and $\Xib$ samples before the final mass selection is shown in 
Fig.~\ref{masses}.
The mass resolution of the $\Xi$ and backgrounds of the samples agreed well.
Only events with the $\Lambda\pi$ invariant mass between 1.309 GeV/$c^2$ 
and 1.333 GeV/$c^2$ were used for analysis.

\begin{figure}[hbt]
\centerline{\psfig{figure=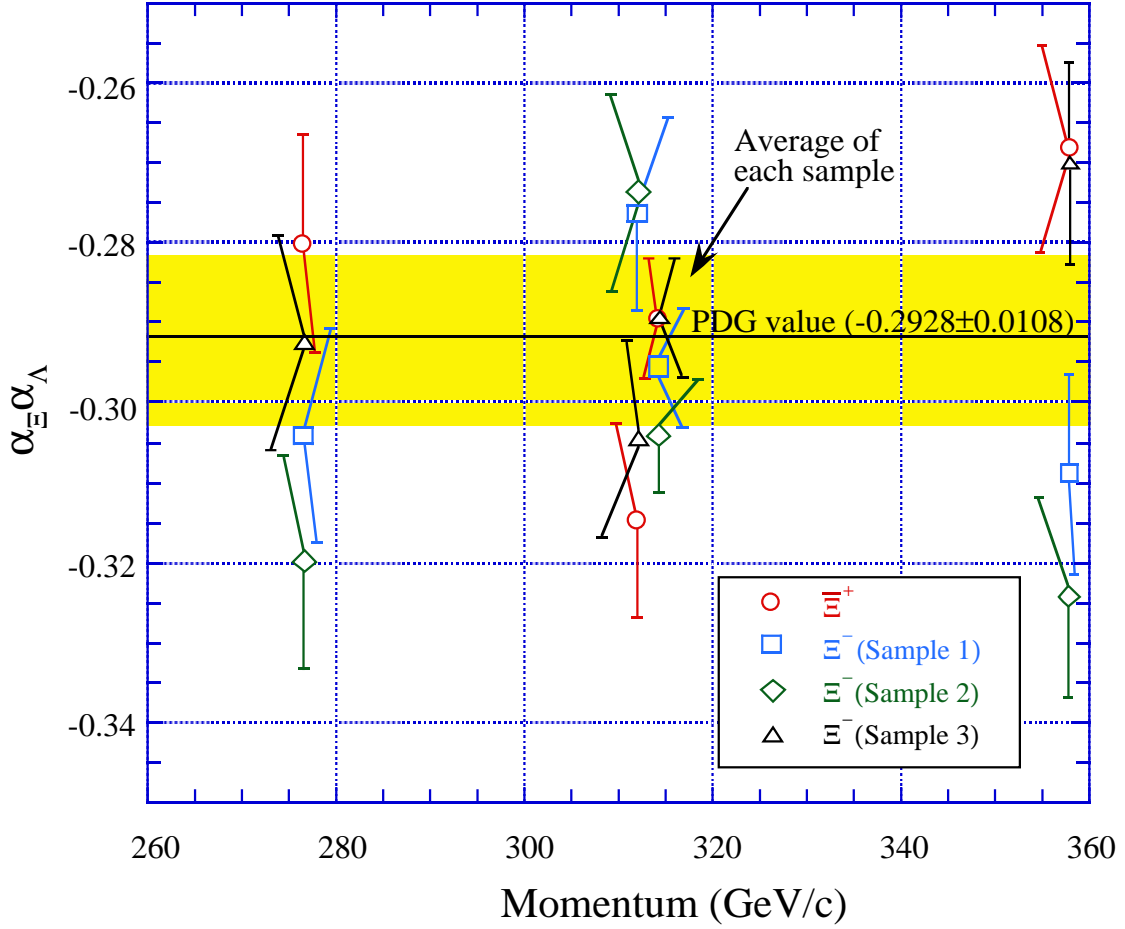,width=6in}}
\caption{Results on $\alpha_{\Xi}\alpha_{\Lambda}$ as a function of
the momentum of the $\Xi$. The shaded area is a one-standard-deviation band
centered at the world average.}
\label{hmc}
\end{figure}

Two different methods were applied to measure $A_{\Xi\Lambda}$.
In the first approach, the acceptance that affected the cos$\theta_{p\Lambda}$ 
distribution given in Eq.~(\ref{eq:helicity}) was 
determined with the Hybrid Monte Carlo (HMC) method\cite{GB2}
before the value of $\alpha_{\Xi}\alpha_{\Lambda}$ was calculated  
for the $\Xi^{-}$ and the $\overline{\Xi}^{+}$ samples separately.
For each event, up to 200 HMC events were generated with a 
uniform distribution in cos$\theta_{p\Lambda}$, but the rest of the 
kinematic quantities such as decay vertices and the momentum of the 
$\Lambda$ were taken from the data. 
The event was included in the asymmetry measurement when 10 of the 
generated HMC events satisified all the requirements in the software that 
simulated the geometry of the spectrometer and the triggers. 
Based on about 70,000 $\overline{\Xi}^{+}$ decays $\alpha_{\overline{\Xi}}
\alpha_{\overline{\Lambda}}$ was found to be $-0.2894 \pm 0.0073$.
Three independent $\Xi^{-}$ samples, each with approximately
70,000 events, were selected from a larger pool of events in such a way that 
the resulting momentum distribution 
of each $\Xi^-$ sample was identical to that of the $\overline{\Xi}^+$. 
In doing so the difference in the momentum-dependent 
acceptance between the $\Xi^{-}$ 
and $\overline{\Xi}^{+}$ samples was reduced. 
The values of $\alpha_{\Xi}\alpha_{\Lambda}$ for these data sets were 
determined to 
be $-0.2955 \pm 0.0073$, $-0.3041 \pm 0.0073$, and $-0.2894 \pm 0.0073$,
giving an average of $-0.2963 \pm 0.0042$.
As shown in Fig.~\ref{hmc}, these results are in good agreement
with the world
average\cite{PDG} and were stable with respect to the momentum of the $\Xi$.
This method gave a value of $0.012 \pm 0.014$ for $A_{\Xi\Lambda}$.
The systematic uncertainty, estimated by varying some of the event-
selection requirements, was insignificant.

In the second approach, the difference in $\alpha_{\Xi}\alpha_{\Lambda}$
between $\Xi^-$ and $\overline{\Xi}^+$ was
determined directly without unfolding the acceptance in cos$\theta_{p\Lambda}$.
Two data sets can be compared by defining 
\vspace{5pt}
\begin{equation} \label{eq:ratio}
R(cos \theta_{p\Lambda}) =
\frac{\epsilon_1(cos \theta_{p\Lambda})}
{\epsilon_2(cos \theta_{p\Lambda})}
\frac{[ 1 + (\alpha_{\Lambda} \alpha_{\Xi})_1 cos \theta_{p\Lambda}]}
{[ 1 + (\alpha_{\Lambda} \alpha_{\Xi})_2 cos \theta_{p\Lambda}]}
\vspace{5pt}
\end{equation}
where $R(cos \theta_{p\Lambda})$ is the ratio of the probabilities of
getting cos$\theta_{p\Lambda}$ in the two samples,
and the $\epsilon$'s are the acceptance functions of the cos$\theta_{p\Lambda}$
distributions.

When two sets of $\Xi^-$ events are compared, $R$ is a measure of how
well the acceptances agree.
Without any corrections,
detailed studies showed that the acceptance in cos$\theta_{p\Lambda}$
was strongly dependent
on the momentum of the $\Xi^-$, but was insensitive to the polarization of the
$\Xi^-$ or other variations in the experiment, and that $R$ was consistent 
with unity and independent of cos$\theta_{p\Lambda}$ down to a few 
$\times~10^{-3}$ level\cite{e871}.
This unique feature is due to the fact that the unit vector 
$\hat{\mathbf{p}}_{\Lam}$ defining the helicity frame
changes from event to event over the entire phase space in 
the rest frame of the $\Xi$.
Any systematic bias due to local inefficiencies of the experiment in the 
laboratory is mapped into a wide range of cos$\theta_{p\Lambda}$ and 
thus highly diluted.

\begin{figure}[hbt]
\centerline{\psfig{figure=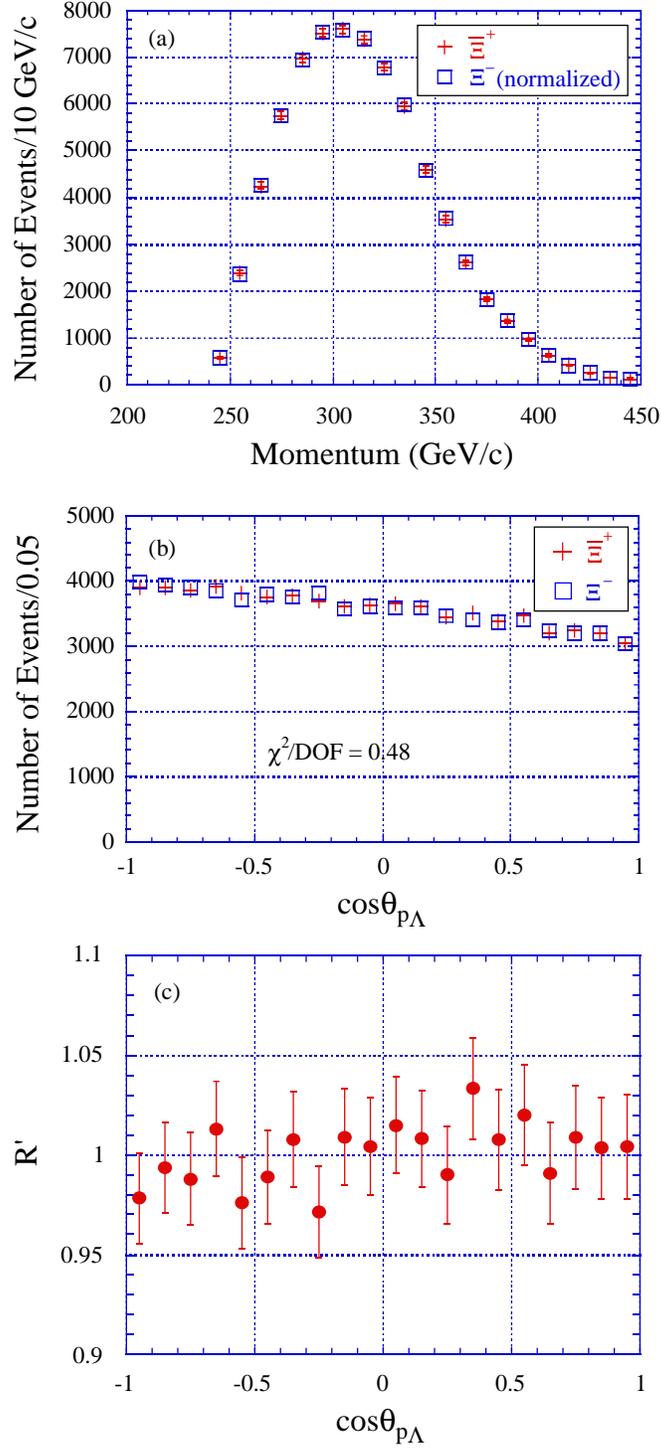,width=3.5in}}
\caption{Comparison of $\Xib$ and $\Xi^-$ events after the momentum 
distributions are normalized. The momentum distributions are shown in (a). 
The cos$\theta_{p\Lambda}$ distributions are shown in (b). $R'$ as a function 
of cos$\theta_{p\Lambda}$ is shown in (c).
}
\label{compare}
\end{figure}

In the study of CP symmetry, a sample of $\Xi^-$ events was selected in such
a way that
the resulting $\Xi^-$ momentum spectrum was identical to that of the
$\overline{\Xi}^+$ sample.
This removed any difference in the momentum spectra due to the 
different mechanism for producing
particles and anti-particles by protons, and
ensured that $\epsilon(cos \theta_{p\Lambda})$ was identical for
both data sets.
In this case, Eq.~(\ref{eq:ratio}) is simply 
\vspace{5pt}
\begin{equation} \label{eq:ratio1}
R'(cos \theta_{p\Lambda}) =
\frac{ 1 + \alpha_{\Lam} \alpha_{\Xi} cos \theta_{p\Lambda}}
{ 1 + \alpha_{\overline{\Lam}} \alpha_{\overline{\Xi}} cos \theta_{p\Lambda}}
= \frac{ 1 + \alpha_{\Lam} \alpha_{\Xi} cos \theta_{p\Lambda}}
{ 1 + (\alpha_{\Lam} \alpha_{\Xi} - D) cos \theta_{p\Lambda}}
\vspace{5pt}
\end{equation}
where $\alpha_{\Lam}\alpha_{\Xi}$ is taken to be $-0.2928$\cite{PDG}, and 
$D$ = $\alpha_{\Lam}\alpha_{\Xi} - \alpha_{\overline{\Lam}}
\alpha_{\overline{\Xi}}$ can be determined by fitting $R'$ as a 
function of cos$\theta_{p\Lambda}$.
With approximately 70,000 $\Xi^-$ events along with equal number of
$\overline{\Xi}^+$ decays, the comparison of the momentum and 
cos$\theta_{p\Lambda}$ distributions of the $\Xi$ samples, 
and the resulting $R'$ are shown in Fig.~\ref{compare}.
The $\theta_{p\Lambda}$ distributions agree well, with a 
$\chi^{2}$ per degree of freedom of 0.45.
$D$ was found to be $-0.011 \pm 0.009$.
This implied that $A_{\Xi\Lambda}$ was $0.019 \pm 0.015$, which was
consistent with the result obtained with the HMC method.
As a check, another sample of $\Xi^-$ events was picked to repeat
the measurement, which yielded 
a result of $0.008 \pm 0.015$ for $A_{\Xi\Lambda}$.
Athough this normalization method constituted a powerful cross check for the
HMC approach, it suffered from the fact that we had to rely on the measured 
value of $\alpha_{\Lam}\alpha_{\Xi}$ to determine $A_{\Xi\Lambda}$.
Thus, we preferred to choose the result from the HMC method as our 
measurement for $A_{\Xi\Lambda}$.

In summary, we have searched for direct CP violation in non-leptonic decays of 
charged $\Xi$ and $\Lam$ by determining the asymetry parameter $A_{\Xi\Lam}$.
With approximately 70,000 $\overline{\Xi}^+$ and 210,000 $\Xi^-$ decays, 
we obtained a result of $0.012 \pm 0.014$ for $A_{\Xi\Lam}$.  
Based on the result of $A_\Lambda$ = $-0.013\pm0.022$ from PS185, 
we deduced $A_\Xi$ to be $0.025\pm0.026$.
Our results are consistent with no CP violation at the $10^{-2}$ 
level in the non-leptonic decays of charged $\Xi$ and $\Lam$.

We would like to thank the staffs of Fermilab for their excellent 
support.
This work was supported in part by the National Science Foundation, 
and the Director, Office of Science, Office of High
Energy and Nuclear Physics, of the U.S. Department of Energy under Contract No.
DE-AC03-76SF00098.

\pagebreak

\end{document}